\documentclass{iaus}
\usepackage{graphicx}

\title[Cosmological effects on the GRB fluxes] %% give here short title %%
{Cosmological effects on the observed flux and\\
fluence distributions of gamma-ray bursts}

\author[\v{R}\'{\i}pa J., M\'esz\'aros A., Ryde F.]   %% give here short author list %%
{Jakub \v{R}\'{\i}pa$^1$, Attila M\'esz\'aros$^2$ and Felix Ryde$^{3}$}

\affiliation{$^1$Institute for the Early Universe, Ewha Womans University, 120-750 Seoul, Korea \\
email: {\tt ripa@ewha.ac.kr; ripa@sirrah.troja.mff.cuni.cz} \\[\affilskip]
$^2$Astronomical Institute, Charles University, 180 00 Prague, Czech Republic \\
email: {\tt meszaros@cesnet.cz} \\[\affilskip]
$^3$ Department of Physics, Royal Institute of Technology, AlbaNova University Center, \\SE-106 91 Stockholm, Sweden\\
email: {\tt felix@particle.kth.se}}

\pubyear{2012}
\volume{279}  %% insert here IAU Symposium No.
\pagerange{385--386}
\jname{Death of Massive Stars: Supernovae \& Gamma-Ray Bursts}
\editors{A.C. Editor, B.D. Editor \& C.E. Editor, eds.}
\doi{10.1017/S1743921312013464}
\begin{document}

\maketitle

\begin{abstract}
Several claims have been put forward that an essential fraction of long-duration BATSE gamma-ray bursts should lie at redshifts larger than 5. This point-of-view follows from the natural assumption that fainter objects should, on average, lie at larger redshifts. However, redshifts larger than 5 are rare for bursts observed by Swift. The purpose of this article is to show that the most distant bursts in general need not be the faintest ones. We derive the cosmological relationships between the observed and emitted quantities, and arrive at a prediction that is tested on the ensembles of BATSE, Swift and Fermi bursts. This analysis is independent on the assumed cosmology, on the observational biases, as well as on any gamma-ray burst model. We arrive to the conclusion that apparently fainter bursts need not, in general, lie at large redshifts. Such a behaviour is possible, when the luminosities (or emitted energies) in a sample of bursts increase more than the dimming of the observed values with redshift. In such a case dP(z)/dz $>$ 0 can hold, where P(z) is either the peak-flux or the fluence. This also means that the hundreds of faint, long-duration BATSE bursts need not lie at high redshifts, and that the observed redshift distribution of long Swift bursts might actually represent the actual distribution.
\end{abstract}

\firstsection % if your document starts with a section,
\section{Introduction}
This article briefly summarizes works published by \cite{mesz11-aa} and \cite{mesz11-ap}.
It can be shown that for the observed peak-flux or fluence $P(z)$ of gamma-ray bursts (GRBs) it holds:
$
P(z) = \frac{(1+z)^N \tilde{L}(z)}{4\pi d_\mathrm{l}^2(z)},
$
$z$ is the redshift; $d_\mathrm{l}(z)$ is the luminosity distance;
$\tilde{L}(z)$ is isotropic peak-luminosity or emitted energy.
$N=0;\, 1;\, 2$ depending on the units of the observables:
$N=0$ if $P(z)$ is peak-flux in units erg/(cm$^2$s) and then $\tilde{L}(z)$ in units erg/s;
$N=1$ if $P(z)$ is fluence in units erg/cm$^2$ and then $\tilde{L}(z)$ in ergs or
$N=1$ if $P(z)$ is peak-flux in units ph/(cm$^2$s) and then $\tilde{L}(z)$ in units ph/s;
$N=2$ if $P(z)$ is fluence in units ph/cm$^2$ and then $\tilde{L}(z)$ is in photons.
$P(z)$ is given by photons with  energies $E_1 \leq E \leq E_2$, where
$E_{1,2}$ is the detector range. $\tilde{L}_\mathrm{ph}(z)$ is from
energy range $E_1(1+z) \leq E \leq E_2(1+z)$.

It is a standard cosmology that for small redshifts ($z \ll 0.1$) $d_\mathrm{l}(z)\propto z$, and for high redshifts
$\lim_{z \rightarrow \infty} \frac{d_\mathrm{l}(z)}{1+z} =$ finite and positive number
for any $H_o$, $\Omega_M$, $\Omega_{\Lambda}$.
If we assume that $\tilde{L}(z) \propto (1+z)^{q}; N+q>2$ for $z \rightarrow \infty$
then $\frac{dP(z)}{dz} > 0$, for $z \rightarrow \infty$.
If $\tilde{L}(z)$ increases with $z$ faster than $\propto (1+z)^{2-N}$ for high redshifts,
then inverse behaviour can happen, and the apparently brighter GRBs can be at higher redshifts; $dP(z)/dz > 0$
can occur.

For small redshifts $z$, $P(z)$ decreases as $z^{-2}$, but if
$ \tilde{L}(z) \propto (1+z)^{q}$ and $N+q > 2$
then $P(z)$ starts to increase as $z^{N+q-2}$ for high $z$.
To answer the question where the $z_\mathrm{turn}$ is, i.e. where $dP(z)/dz = 0$, one can
simply search for the minimum of $Q(z) = (1+z)^{N+q}/d_\mathrm{l}(z)^2$, i.e., when $dQ(z)/dz = 0$.

\section{Samples of long GRBs}

\begin{figure}
\begin{center}
\includegraphics[width= {0.95\columnwidth}]{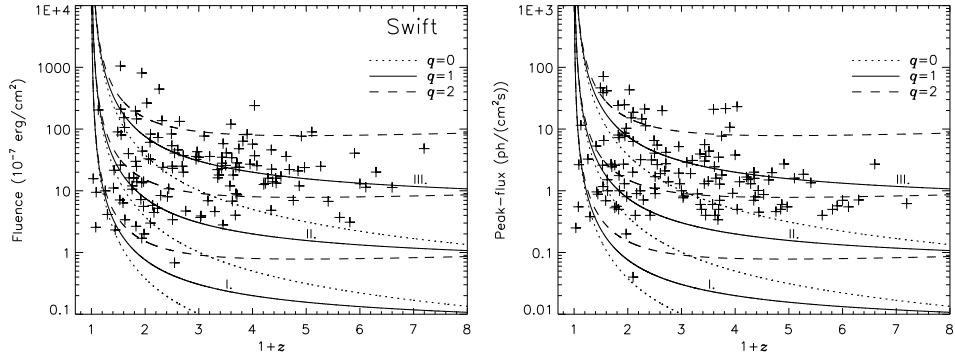}
\caption{Distributions of fluences (left panel) and peak-fluxes (right panel) Swift long GRBs.
On the left panel the curves denote the values of fluences for $\tilde{E}_\mathrm{iso} = \tilde{E_o}
(1+z)^q $ (constants $\tilde{E_o}$ are in $10^{51}$\,erg: I. 0.1; II. 1.0;
III.  10.0). On the right panel the curves denote the values of peak-fluxes for $\tilde{L}_\mathrm{iso} =
\tilde{L_o} (1+z)^q$ (constants $\tilde{L_o}$ are in $10^{58}$\,ph/s: I.
0.01; II. 0.1; III. 1.0). Here $N=1$ thus value $q=1$ is the limiting case.}
\label{fig1}
\end{center}
\end{figure}

We studied 134 long Swift GRBs with know redshifts. The distribution of observed flunces/peak fluxes together with
the curves of constant $\tilde{E}_\mathrm{iso}$ and $\tilde{L}_\mathrm{iso}$ are shown in Fig.\,\ref{fig1}.
We also studied 6 Fermi GRBs, 9 BATSE GRBs with measured redshifts and 13 BATSE GRBs with calculated pseudo-redshifts.
These samples also demonstrate that fainter bursts can well have smaller redshifts (for details see \cite{mesz11-aa}).

\section{Conclusion}

The theoretical study of the $z$-dependence of the observed fluences and peak-fluxes of
GRBs have shown that fainter bursts could well have smaller redshifts.
This is really fulfilled for the four different samples of long GRBs.
These results do not depend on the cosmological parameters and on GRB models.

\acknowledgements
This study was supported by the OTKA grant K77795, by the Grant Agency of the Czech Republic grants No. P209/10/0734, by the Research Program MSM0021620860 of the Ministry of Education of the Czech Republic, and by Creative Research Initiatives (RCMST) of MEST/NRF and the World Class University grant no R32-2008-000-101300.

\end{document}